\documentclass[a4paper,twocolumn]{esapub2005} 
\usepackage{graphicx}
\usepackage{times}
\usepackage{natbib}

\newcommand{\solar}{\mbox{$_{\normalsize\odot}$}}

\newcommand{\alp}{\mbox{$\alpha$}}
\newcommand{\farcm}{\hbox{$.\mkern-4mu^\prime$}}
\newcommand{\lsim}{\ \raise
-2.truept\hbox{\rlap{\hbox{$\sim$}}\raise5.truept\hbox{$<$}\ }}
\newcommand{\gsim}{\ \raise
-2.truept\hbox{\rlap{\hbox{$\sim$}}\raise5.truept\hbox{$>$}\ }}
\newcommand{\simsim}{\ \raise
-1.5truept\hbox{\rlap{\hbox{$\sim$}}\raise3.5truept\hbox{$\sim$}\ }}

\long\def\symbolfootnote[#1]#2{\begingroup%
\def\thefootnote{\fnsymbol{footnote}}\footnote[#1]{#2}\endgroup}

\title{A {\em Hubble} View of Star Forming Regions in the Magellanic
Clouds}

\author[1]{Dimitrios A. Gouliermis\footnote{Research supported by the
German Research Foundation (Deutsche Forschungsgemeinschaft -- DFG), and
the German Aerospace Center (Deutsche Zentrum f\"{u}r Luft und Raumfahrt
-- DLR).}} 

\author[1]{Thomas K. Henning}
\author[1]{Wolfgang Brandner}
\author[2]{Michael R. Rosa} 
\author[3]{Andrew E. Dolphin}
\author[1]{Markus Schmalzl}
\author[1]{Eva Hennekemper}
\author[4]{Hans Zinnecker} 
\author[5]{Nino Panagia}
\author[6]{You-Hua Chu}
\author[7]{Bernhard Brandl}
\author[1]{Sascha P.~Quanz}
\author[5]{Massimo Robberto}
\author[8]{Guido De Marchi} 
\author[6]{Robert A. Gruendl}
\author[2]{Martino Romaniello}

\affil[1]{Max-Planck-Institut f\"ur Astronomie, K\"onigstuhl 17, 69117
Heidelberg, Germany}
\affil[2]{European Southern Observatory, Karl-Schwarzschild-Strasse 2,
85748 Garching, Germany}
\affil[3]{Raytheon Corporation, USA}
\affil[4]{Astrophysikalisches Institut Potsdam, An
der Sternwarte 16, 14482 Potsdam, Germany}
\affil[5]{Space Telescope Science Institute, 3700 San Martin Drive,
Baltimore, MD 21218, USA}
\affil[6]{Department of Astronomy, University of Illinois, 1002 West Green
Street, Urbana, IL 61801, USA}
\affil[7]{Sterrewacht Leiden, P.O. Box 9513, Niels Bohrweg 2, 2300 RA
Leiden, The Netherlands}
\affil[8]{ESA, Space Science Department, Keplerlaan 1, 2200 AG Noordwijk,
The Netherlands}

\begin{document}


\maketitle

\begin{abstract}

The Magellanic Clouds (MCs) offer an outstanding variety of young stellar 
associations, in which large samples of low-mass stars (with 
$M$~\lsim~1~M\solar) currently in the act of formation can be resolved and 
explored sufficiently with the {\em Hubble} Space Telescope. These 
pre-main sequence (PMS) stars provide a unique snapshot of the star 
formation process, as it is being recorded for the last 20 Myr, and they 
give important information on the low-mass Initial Mass Function (IMF) of 
their host environments. We present the latest results from observations 
with the {\em Advanced Camera for Surveys} (ACS) of such star-forming 
regions in the MCs, and discuss the importance of {\em Hubble} for a 
comprehensive collection of substantial information on the most recent 
low-mass star formation and the low-mass IMF in the MCs.

\end{abstract}

\section{Introduction}

The Large and Small Magellanic Cloud (LMC, SMC) are the closest undisrupted 
neighboring dwarf galaxies to our own. They have four to five times lower 
metallicities than the Milky Way (MW), while their gas-to-dust ratio is much 
higher, forming environments resembling those of the early universe. The 
Magellanic Clouds (MCs) show clear evidence for energetic star formation 
activity with {\sc H~i} shells \citep{1980MNRAS.192..365M}, 
\citep{1999AJ....118.2797K}, {\sc H~ii} regions \citep{1956ApJS....2..315H}, 
\citep{1976MmRAS..81...89D}, and molecular clouds 
\citep{1999PASJ...51..745F}, \citep{2002ApJ...566..857T}, all linked to 
ongoing star formation, as it is observed in young stellar systems, the {\em 
Stellar Associations} \citep{1970AJ.....75..171L}, 
\citep{1999AJ....117..238B}, \citep{2003A&A...405..111G}. Both MCs contain a 
variety of such stellar systems, the age and IMF of which become very 
important sources of information on their recent star formation. They 
provide a rich sample of targets for the comprehensive study of current star 
formation in low-metallicity environments. Considering that the MCs are so 
close to us ($\sim$ 50 kpc and 60 kpc), they are indeed ideal laboratories 
for a detailed study of clustered star formation and the IMF in the early 
universe, and {\em Hubble}'s contribution is fundamental in such a study.

\begin{figure}[t]
\centering
\includegraphics[width=0.95\linewidth]{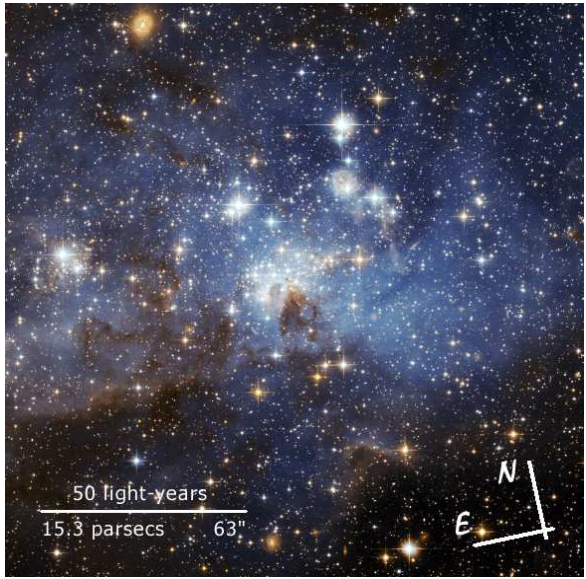}
\caption{Color-composite image from ACS/WFC observations in the filters
$F555W$ and $F814W$ ($V$- and $I$-equivalent) of the LMC star-forming
region LH~95/N~64. This sharp image, presented at the 2006 General
Assembly of the International Astronomical Union, reveals a large number
of low-mass infant stars coexisting with young massive ones. These
observations, being the deepest ever taken towards the LMC, allow us to
explore the scientific gain that can be achieved for MCs studies using
high spatial resolution photometry from {\em Hubble}.  Image credit:
NASA, ESA and D. A. Gouliermis (MPIA). Acknowledgments: Davide de
Martin (ESA/Hubble).\label{fig-ima}}
\end{figure}

\begin{figure}[t]
\centering
\includegraphics[width=1.\linewidth]{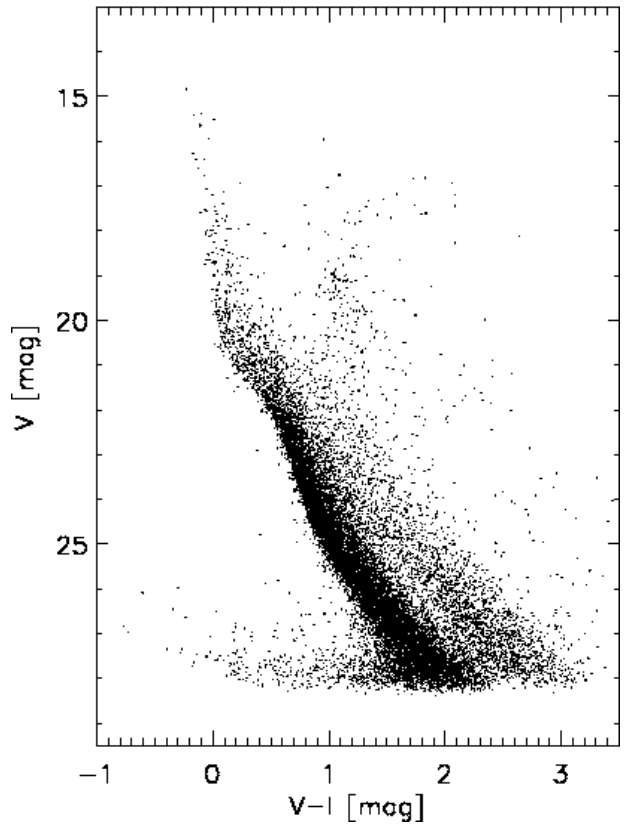}
\caption{The $V-I$, $V$ CMD of the stars detected with ACS/WFC in the
region of LH~95/N~64. These observations, with a detection limit of
$V\simeq$~28.4 mag ($M$~\lsim~0.5~M{\solar}), reveal a unique sample of
$\sim$~2,450 PMS stars, easily distinguished as a secondary red
sequence, almost parallel to the faint part of the main sequence. These
stars are found to be concentrated in the central part of the
association and in surrounding compact clusters, and their spatial
distribution is in excellent coincidence with the loci of the brightest
MS stars \citep{Gouliermis2007b}.\label{fig-cmd}}
\end{figure}

\section{A New View of MCs Associations}

Stellar associations contain the richest sample of young bright stars in a 
galaxy. Consequently our knowledge on the young massive stars of the MCs has 
been collected from photometric and spectroscopic studies of young stellar 
associations \citep{2006lgal.symp..164M}. However, the picture of these 
stellar systems changed when {\em Hubble} observations revealed that MCs 
associations are not mere aggregates of young bright stars alone, but they 
also host large numbers of faint PMS stars \citep{2006ApJ...636L.133G}, 
\citep{2006ApJ...640L..29N}. Although nearby galactic OB associations are 
known to be significant hosts of such stars \citep{2002AJ....124..404P}, 
\citep{2004AJ....128.2316S}, \citep{2007prpl.conf..345B}, PMS studies in the 
MCs with {\em Hubble} were focused only on the surrounding field of the 
supernova 1987A \citep{2000ApJ...539..197P}, cluster NGC~1850 
\citep{1994ApJ...435L..43G}, and the star-burst of 30 Doradus 
\citep{2001AJ....122..858B}, \citep{2006A&A...446..955R} all in the LMC. 
However, these studies are limited by crowding, even at the angular 
resolution facilitated by {\em Hubble}.

To learn more about low-mass PMS stars in the MCs, one has to study less
crowded regions like young stellar associations. Indeed, an
investigation on the main-sequence IMF of the LMC association LH~52 with
HST/WFPC2 observations by \citep{2005ApJ...623..846G} revealed $\sim$
500 low-mass candidate PMS stars easily distinguishable in the $V-I$,
$V$ Color-Magnitude Diagram (CMD) \citep{2006ApJ...636L.133G}. More
recently, deeper observations with the Wide-Field Channel (WFC) of ACS
of another LMC association (the star-forming region LH~95/N~64) revealed
the coexistence of PMS stars and early-type stars in such stellar
systems (Figure \ref{fig-ima}).

These one-of-a-kind observations dramatically changed the picture we had 
for stellar associations in the MCs by revealing a unique rich sample of 
PMS stars in LH~95/N~64 (Figure \ref{fig-cmd}).  The spatial distribution 
of these low-mass members demonstrates the existence of significant 
substructure (``subgroups''), as in the case of galactic OB associations.  
This stellar sub-clustering has its origins possibly in short-lived 
parental molecular clouds within a Giant Molecular Cloud Complex. Each of 
these ``{\em PMS clusters}'' in LH~95/N~64 includes a few early-type 
stars. Such stars have been identified as candidate Herbig Ae/Be (HAeBe) 
stars due to their strong H\alp\ emission \citep{2002A&A...381..862G}. 
Near-IR spectroscopic study with VLT/SINFONI (ESO Program 078.D-0200) will 
clarify their nature.


\begin{figure}[b!]
\centering
\includegraphics[width=1.0\linewidth]{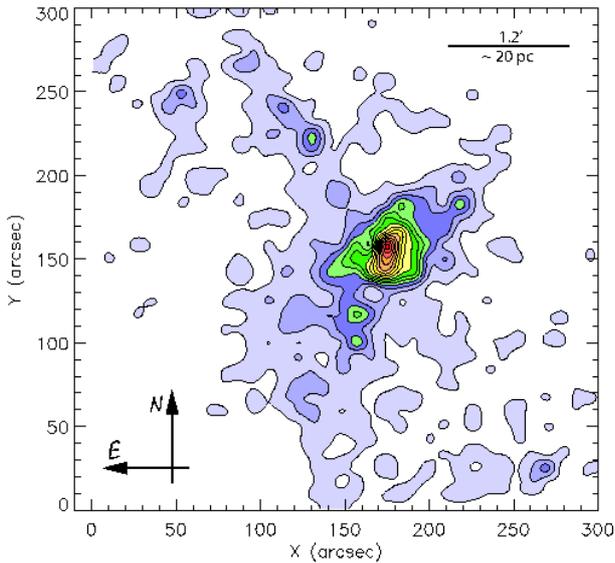}
\caption{Isodensity contour map of the region of NGC~346/N~66 in the SMC
from ACS/WFC observations, constructed from star counts of the PMS stars
\citep{2006ApJS..166..549G}. Isopleths are plotted in steps of
1$\sigma$, $\sigma$ being the standard deviation of the background
surface density. This map demonstrates the existence of statistically
significant concentrations of PMS stars outside the main body of the
association NGC~346 (located at the center).  Such PMS
clusters are suspected to be the product of sequential star formation
triggered by the action of the OB stars in NGC~346, which shape
the southern part, and a supernova, which affects the
northern part of the region \citep{Hennekemper2007}. \label{fig-cmap}}
\end{figure}

\subsection{Stellar Subgroups in MCs Associations}

A spatial behavior similar to the PMS population of LH~95/N~64 is seen in 
PMS stars of the association NGC~346 in the SMC \citep{2006ApJS..166..549G}, 
from observations with ACS/WFC. NGC~346 is located in the brightest {\sc 
H~ii} region of the SMC, N~66, and ACS uncovered the richness of this region 
in PMS stars \citep{2006ApJS..166..549G}, \citep{2006ApJ...640L..29N}. The 
surface density map of the region of NGC~346/N~66 constructed from star 
counts of the PMS stars in the observed field is shown in Figure 
\ref{fig-cmap}. Apart from the association itself (seen as the central large 
concentration) there are at least five distinct concentrations of PMS stars 
with surface stellar density \gsim\ 3$\sigma$ above the background (where 
$\sigma$ is the standard deviation of the background density), which fit the 
description of ``PMS clusters''. The size of each cluster is defined by the 
isopleth corresponding to the local mean density around it, and the 
time-scale within which each PMS cluster was presumably formed is defined by 
their individual CMDs. Although the loci of the PMS stars in the CMD exhibit 
a broadening, which prevents an accurate estimation of their age, it was 
found that the PMS clusters located away from NGC~346 to the north, 
represent the most recent star formation activity in the region 
\citep{Hennekemper2007}.

\begin{figure}[t!]
\centering
\includegraphics[width=.95\linewidth]{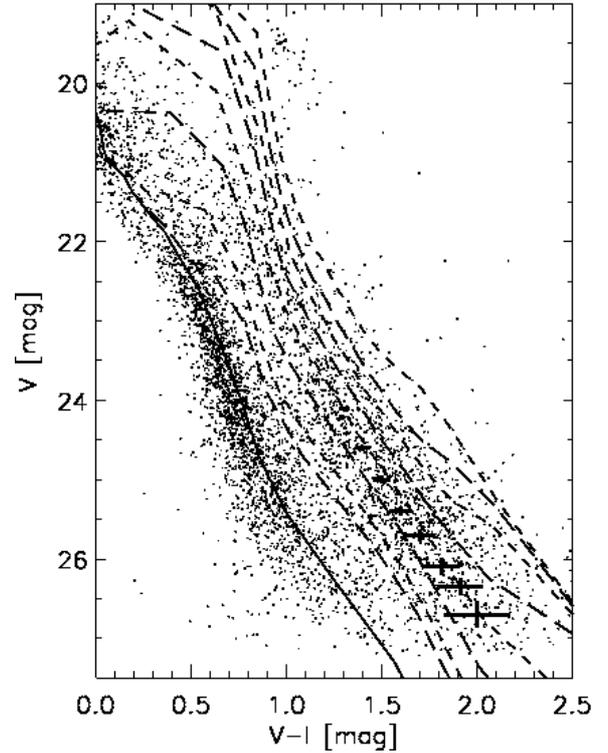}
\caption{Detail of the $V-I$, $V$ CMD of all stars detected with ACS/WFC
imaging in the area of NGC~346 (0\farcm6 around its center). PMS
isochrone models by \citep{2000A&A...358..593S} for ages 0.5 to 15~Myr are
overplotted to demonstrate that the observed broadening of the PMS stars
can be easily misinterpreted as an age-spread. Simulations showed that
this spread can be explained as the result of interstellar reddening of
$E(B-V) \simeq 0.08$~mag alone, or of two star formation events (a true
age-spread) $\sim$5~Myr apart if the reddening is lower. Typical
photometric uncertainties in brightness and color are also shown
\citep{Hennekemper2007}.  \label{fig-pms}}
\end{figure}

\subsection{CMD Broadening of PMS Stars}

The loci of PMS stars in the CMD often show a widening, which could be
evidence for an age-spread \citep{2000ApJ...540..255P}.  The low-mass
population in subgroups within OB associations of the MW exhibits little
evidence for significant age-spreads on time-scales \gsim~10~Myr
\citep{2007prpl.conf..345B}. Although this time-scale is in agreement
with a scenario of rapid star formation and cloud dissipation, age
differences of the order of 10~Myr may be very important for 
understanding of how sequential star formation proceeds. 

Moreover, there are several factors apart from age-spread, such as 
variability and binarity, which can cause considerable deviations of the 
positions of the PMS stars in the CMD \citep{2004AJ....128.2316S}. A 
broadening in the CMD is also observed for the PMS stars of NGC~346/N~66 
(Figure \ref{fig-pms}). Simulations showed that apart from photometric 
uncertainties, binarity and variability, reddening seems to play the most 
important role in the observed widening of the PMS stars, providing false 
evidence for an age-spread if the region suffers from high extinction 
\citep{Hennekemper2007}.

\section{The IMF of MCs Associations}

\begin{figure}[t!]
\centering
\includegraphics[width=0.95\linewidth]{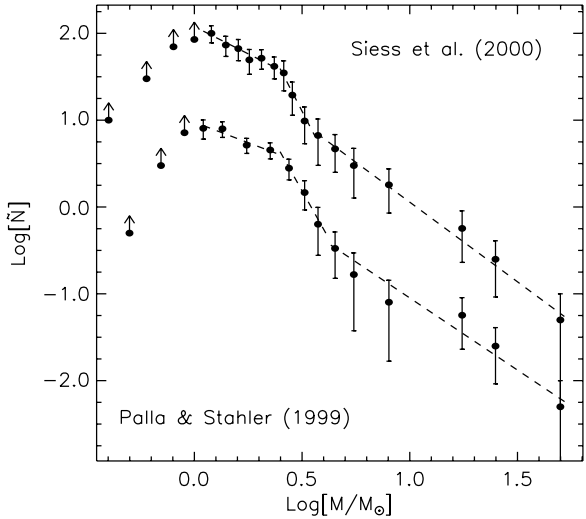}
\caption{The stellar mass spectrum of NGC~602 for the whole mass range
observed with ACS/WFC. The low-mass part (\lsim~6~{\rm M}\solar) was
constructed by counting PMS stars between evolutionary tracks, with the use
of two sets of PMS grids \citep{1999ApJ...525..772P},
\citep{2000A&A...358..593S}. It is found that the IMF seems to be
model-independent and is well represented by a three-part power law
\citep{Schmalzl2007}. \label{fig-imf}}
\end{figure}

Young stellar systems, which host newborn PMS stars, naturally provide the 
testbed for a comprehensive study of the stellar IMF. A coherent sample of 
PMS stars is found with {\em Hubble} from ACS imaging in the vicinity of 
another SMC association, NGC~602, located in the {\sc H~ii} region N~90 
\citep{Schmalzl2007}. The region of NGC~602/N~90 includes no distinct 
subgroups, and therefore, being less complicated than NGC~346/N~66, is 
more suitable for the investigation of the low-mass IMF. For the 
construction, though, of this IMF a mass-luminosity relation derived from 
evolutionary models cannot be used due to the spread of the PMS stars, 
also apparent in the CMD of NGC~602. Instead, counting the PMS stars 
between evolutionary tracks, which represent specific mass ranges, seems 
to be the most accurate method for the construction of their mass spectrum 
(Figure \ref{fig-imf}).

\section{On-going Star Formation in the MCs}

The coexistence of {\sc H~ii} regions and PMS stars in stellar 
associations of the MCs indicate that star formation may be still active 
in their vicinity. Indeed, observations with the {\em Spitzer} Space 
Telescope revealed objects classified as candidate Young Stellar Objects 
(YSOs) in such regions \citep{2006AJ....132.2268M}, 
\citep{2007ApJ...655..212B}, and {\em Hubble}'s contribution has been very 
important in disentangling their nature \citep{2005ApJ...634L.189C}. The 
region of NGC~602/N~90 is also found with {\em Spitzer} to host possible 
YSOs, and the comparison of the loci of these IR-bright sources with the 
{\em Hubble} images interestingly showed PMS stars to be their optical 
counterparts \citep{Gouliermis2007}. A variety of objects is discovered to 
coincide with these candidate YSOs, such as single highly embedded 
sources, small compact PMS clusters, as well as features similar to 
``Elephant Trunks'', all located at the periphery of NGC~602, along the 
dust ridges of the molecular cloud presumably blown-away by the action of 
the association itself \citep{Gouliermis2007}.


\section*{Acknowledgments}

D. A. Gouliermis kindly acknowledges the support of the German Research
Foundation through the individual grant 1659/1-1. Based on observations
made with the NASA/ESA Hubble Space Telescope, obtained from the data
archive at the Space Telescope Science Institute. STScI is operated by
the Association of Universities for Research in Astronomy, Inc. under
NASA contract NAS 5-26555.


\end{document}